\documentclass[preprints,article,accept,moreauthors,pdftex]{Definitions/mdpi}

\pdfoutput=1

\firstpage{1}
\pubvolume{9}
\issuenum{15}
\articlenumber{1794}
\pubyear{2021}
\copyrightyear{2021}
\externaleditor{Academic Editor: Vicente Coll-Serrano}
\datereceived{26 June 2021}
\dateaccepted{26 July 2021}
\datepublished{28 July 2021}
\hreflink{https://\linebreak doi.org/10.3390/math9151794} % If needed use \linebreak
\Title{Multi-Transformer: A New Neural Network-Based Architecture for Forecasting S\&P Volatility}

\TitleCitation{Multi-Transformer: A New Neural Network-Based Architecture for Forecasting S\&P Volatility}

\Author{{Eduardo Ramos-P\'erez} 
$^{1}$, Pablo J. Alonso-Gonz\'alez $^{2,}$* and Jos\'e Javier N\'u\~nez-Vel\'azquez {$^{2}$}}

\AuthorNames{Eduardo Ramos-P\'erez, Pablo J. Alonso-Gonz\'alez and Jos\'e Javier N\'u\~nez-Vel\'azquez}

\AuthorCitation{Ramos-P\'erez, E.; Alonso-Gonz\'alez, P.J.; N\'u\~nez-Vel\'azquez, J.J.}
\address{%
$^{1}$ \quad {Faculty of Economics,} Universidad de Alcal\'a, Plaza de la Victoria 2, 28802 Alcal\'a de Henares, Madrid, Spain; ramos.perez.e@gmail.com\\
$^{2}$ \quad Economics Department, Universidad de Alcal\'a, Plaza de la Victoria 2, \mbox{28802 Alcal\'a de Henares, Madrid, Spain};  josej.nunez@uah.es\\
}
\corres{Correspondence: pablo.alonsog@uah.es}

\abstract{Events such as the Financial Crisis of 2007--2008 or the COVID-19 pandemic caused significant losses to banks and insurance entities. They also demonstrated the importance of using accurate equity risk models and having a risk management function able to implement effective hedging strategies. Stock volatility forecasts play a key role in the estimation of equity risk and, thus, in the management actions carried out by financial institutions. Therefore, this paper has the aim of proposing more accurate stock volatility models based on novel machine and deep learning techniques. This paper introduces a neural network-based architecture, called Multi-Transformer. Multi-Transformer is a variant of Transformer models, which have already been successfully applied in the field of natural language processing. Indeed, this paper also adapts traditional Transformer layers in order to be used in volatility forecasting models. The empirical results obtained in this paper suggest that the hybrid models based on Multi-Transformer and Transformer layers are more accurate and, hence, they lead to more appropriate risk measures than other autoregressive algorithms or hybrid models based on feed forward layers or long short term memory cells.}

\keyword{deep learning; neural networks; risk management; stock volatility; transformer}

\begin{document}
%%%%%%%%%%%%%%%%%%%%%%%%%%%%%%%%%%%%%%%%%%
%\setcounter{section}{-1} %% Remove this when starting to work on the template.
%\section{How to Use this Template}

%The template details the sections that can be used in a manuscript. Note that the order and names of article sections may differ from the requirements of the journal (e.g., the positioning of the Materials and Methods section). Please check the instructions on the authors' page of the journal to verify the correct order and names. For any questions, please contact the editorial office of the journal or support@mdpi.com. For LaTeX-related questions please contact latex@mdpi.com.
%The order of the section titles is: Introduction, Materials and Methods, Results, Discussion, Conclusions for these journals: aerospace,algorithms,antibodies,antioxidants,atmosphere,axioms,biomedicines,carbon,crystals,designs,diagnostics,environments,fermentation,fluids,forests,fractalfract,informatics,information,inventions,jfmk,jrfm,lubricants,neonatalscreening,neuroglia,particles,pharmaceutics,polymers,processes,technologies,viruses,vision

\section{Introduction}
\label{VolII_Intro}
Since the Financial Crisis of 2007--2008, financial institutions have enhanced their risk management framework in order to meet the new regulatory requirements set by Solvency II or Basel III. These regulations have the aim of measuring the risk profile of financial institutions and minimizing losses from unexpected events such as the European sovereign debt crisis or COVID-19 pandemic. Even though banks and insurance entities have reduced their losses thanks to the efforts made in the last years, unexpected events still cause remarkable losses to financial institutions. Thus, efforts are still required to further enhance market and equity risk models in which stock volatility forecasts play a fundamental role. Volatility, understood as a measure of an asset uncertainty \cite{hul_2015,rr_2015}, is not directly observed in stock markets. Thus, taking into consideration the stock market movements, a~statistical model is applied in order to compute the volatility of a~security.

GARCH-based models \cite{eng_1982,Bollerslev_1986} are widely used for volatility forecasting purposes. This family of models is especially relevant because it takes into consideration the volatility clustering observed by~\cite{man_1963}. Nevertheless, as~the persistence of conditional variance tends to be close to zero, {Refs.}~\cite{el_1999,hmp_2004a,hmp_2004b,hp_2012} developed more flexible variations of the traditional GARCH models. In~addition, the~models introduced by~\cite{Nelson_1991} (EGARCH) and~\cite{gjr_1993} (GJR-GARCH) take into consideration that stocks volatility behaves differently depending on the market trend, bearish or bullish. Multivariate GARCH models were developed by~\cite{ke_1982,egk_1984}. \citet{bew_1988} {applied} %MDPI:  It's not allowed to start a sentence with a reference number, we added author name, please confrim.
the previous model to financial time series, while~\cite{tt_2002} introduced a time-varying multivariate GARCH. Dynamic conditional correlation GARCH, BEKK-GARCH and Factor-GARCH were other variants of this family that were developed by~\cite{eng_2002,ek_1995,enr_1990}, respectively. Finally, it is worth mentioning that, in~contrast to classical GARCH, the~first-order zero-drift GARCH model (ZD-GARCH) proposed by~\cite{zzl_2018} is non-stationary regardless of the sign of Lyapunov exponent and, thus, it can be used for studying heteroscedasticity and conditional heteroscedasticity~together.

Another relevant family is composed by stochastic volatility models. As~they assume that volatility follows its own stochastic process, these models are widely used in combination with Black--Scholes formula to assess derivatives price. The~most popular process of this family is the~\cite{Heston_1993} model which assumes that volatility follows an Cox-Ingersoll-Ross~\cite{CIR_1985} process and stock returns a Brownian motion. The~main challenge of the Heston model is the estimation of its parameters. {Refs.}~%MDPI: We added refs., please confirm.
\cite{mt_1990,as_1999} proposed a generalized method of moments to obtain the parameters of the stochastic process, while~\cite{dk_1997,br_2004,dan_2004,and_2009} used a simulation approach to estimate them. Other relevant stochastic volatility processes are Hull--White~\cite{HW_1987} and SABR~\cite{HKL_2002} models.

The last relevant family is composed of those models based on machine and deep learning techniques. Even though GARCH models are considered part of the machine learning tool-kit, these models are considered another different family due to the significant importance that they have in the field of stock volatility. Thus, this family takes into consideration the models based on the rest of the machine and deep learning algorithms such as artificial neural networks \cite{Mcculloch_1943}, gradient boosting with regression trees~\cite{Friedman_2000}, random \mbox{forests \cite{Breiman_2001}} or support vector machines \cite{Cortes_1995}. {Refs.}~\cite{gsb_2001,gd_2012,dias_2019} applied machine learning techniques such as Support Vector Machines or hidded Markov models to forecast financial time series.  \citet{hi_2002} {applied} Artificial Neural Networks (ANNs) to demonstrate that the implied volatility forecasted by this algorithm is more accurate than Barone--Adesi and Whaley~models.

ANNs have been also combined with other statistical models with the aim of improving the forecasting power of individual ANNs. The~most common approach applied in the field of stocks volatility is merging GARCH-based models with ANNs. {Refs.}~\cite{Roh_2006,hsz_2012,kfm_2014,m_2014,lu_2016,kw_2018,bk_2018} developed different architectures based in the previous approach for stock volatility forecasting purposes. All these authors demonstrated that hybrid models overcome the performance of traditional GARCH models in the field of stock volatility forecasting. It is also worth mentioning the contribution of~\cite{be_2009}, who combined different GARCH models with ANNs in order to compare their predictive power. ANN-GARCH models have been also applied to forecast other financial time series such as metals \cite{km_2015,kh_2017} or oil \cite{km_2016,Verma_2021} volatility. Apart from the combination with GARCH-based models, ANNs have been merged with other models for volatility forecasting purposes.~\citet{RAN_2019} {merged} ANNs, random forests, support vector machines (SVM) and gradient boosting with regression trees in order to forecast S\&P500 volatility. This model overcame the performance of a hybrid model based on feed forward layers and GARCH.~\citet{VW_2020} {proposed} an architecture based on convolutional neural networks (CNNs) and long-short term memory (LSTM) units to forecast gold volatility. LSTMs were also used by~\cite{JC_2021} to forecast currency exchange rates volatility. It is also worth mentioning that GARCH models have not been only merged with ANNs, \citet{pmc_2018} {combined} SVM with GARCH-based models in order to predict cryptocurrencies~volatility.

The aim of this paper is to introduce a more accurate stock volatility model based on an innovative machine and deep learning technique. For~this purpose, hybrid models based on merging Transformer and Multi-Transformer layers with other approaches such as GARCH-based algorithms or LSTM units are introduced by this paper. Multi-Transformer layers, which are also introduced in this paper, are based on the Transformer architecture developed by~\cite{VSP_2017}. Transformer layers have been successfully implemented in the field of natural language processing (NLP). Indeed, the~models developed by~\cite{BERT_2018,GPT3_2020} demonstrated that Transformer layers are able to overcome the performance of traditional NLP models. Thus, this recently developed architecture is currently considered the state-of-the-art in the field of NLP. In~contrast to LSTM, Transformer layers do not incorporate recurrence in their structure. This novel structure relies on a multi-head attention mechanism and positional embeddings in order to forecast time series. As~\cite{VSP_2017} developed Transformer for NLP purposes, positional embeddings are used in combination with word embeddings. The~problem faced in this paper is the forecasting of stock volatility and, thus, the~word embedding is not needed and the positional embedding has been modified as it is explained in Section~\ref{VolII_Transformer}.

In contrast to Transformer, Multi-Transformer randomly selects different subsets of training data and merges several multi-head attention mechanisms to produce the final output. Following the intuition of bagging, the~aim of this architecture is to improve the stability and accurateness of the attention mechanism. It is worth mentioning that the GARCH-based algorithms used in combination with Transformer and Multi-Transformer layers are GARCH, EGARCH, GJR-GARCH, TrGARCH, FIGARCH and~AVGARCH.

Therefore, three main contributions are provided by this study. First, Transformer layers are adapted in order to forecast stocks volatility. In~addition, an~extension of the previous structure is presented (Multi-Transformer). Second, this paper demonstrates that merging Transformer and Multi-Transformer layers with other models lead to more accurate volatility forecasting models. Third, the~proposed stock volatility models generate appropriate risk measures in low and high volatility regimes. The~Python implementation of the volatility models proposed in this paper is available in this \href{https://github.com/EduardoRamosP/MultiTransformer}{repository}.

As it is shown by the extensive literature included in this section, stock volatility forecasting has been a relevant topic not only for financial institutions and regulators but also for the academia. As~financial markets can suffer drastic sudden drops, it is highly desirable to use models that can adequately forecast volatility. It is also useful to have indicators that can accurately measure risk.This paper makes use of recent deep and machine learning techniques to create more accurate stock volatility models and appropriate equity risk~measures.

The rest of the paper is organized as follows: Section~\ref{VolII_Materials} describes the dataset, the~measures used for validating the volatility forecasts and provides a look at the volatility models used as benchmark. Then, this section presents the volatility forecasting models proposed in this paper, which are based on Transformer and Multi-Transformer layers. As~NLP Transformers need to be adapted in order to be used for volatility forecasting purposes and Multi-Transformer layers are introduced by this paper, explanations about the theoretical background of these structures are also given. The~analysis of empirical results is presented in Section~\ref{VolII_resul}. Finally, the~results are discussed in Section~\ref{VolII_Conclusion}, followed by concluding remarks in Section~\ref{VolII_Conclusion_II}.

\section{Materials and~Methods}
\label{VolII_Materials}
This section is divided in five different subsections. The~first one (Section \ref{VolII_Data}) describes the data for fitting the models. The~measures for validating the accuracy and value at risk (VaR) of each stock volatility model are explained in Section~\ref{VolII_Validation}. Section~\ref{VolII_Benchmark} presents the stock volatility models and algorithms used for benchmarking purposes. Section~\ref{VolII_Transformer} explains the adaptation of Transformer layers in order to be used for volatility forecasting purposes and, finally, the~Multi-Transformer layers and the models based on them are presented in Section~\ref{VolII_MultiTransformer}.

\subsection{Data and Model~Inputs}
\label{VolII_Data}
\begin{figure}[H]
%\begin{center}
\includegraphics[width=0.74\textwidth]{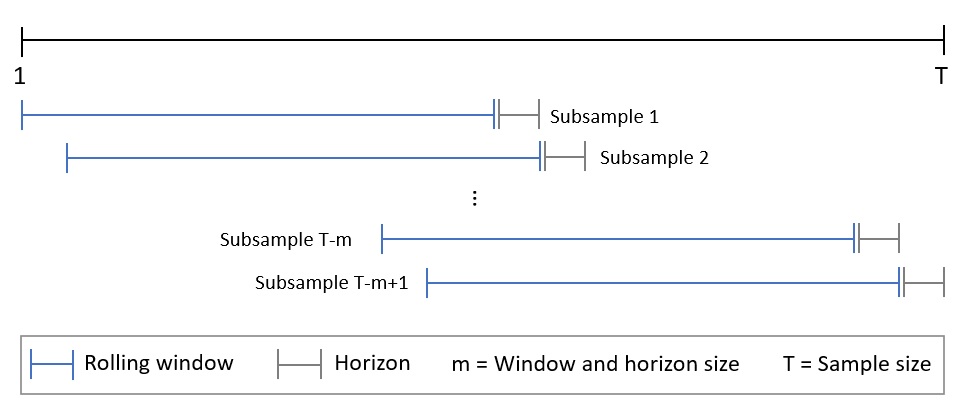}
\caption{Rolling window methodology.}
\label{fig:RollingWindow_Vol_II}
%\end{center}
\end{figure}

The proposed architectures and benchmark models are fitted using the rolling window approach (see Figure~\ref{fig:RollingWindow_Vol_II}). This widely used methodology has been applied in finance, among~others, by~\cite{S_1998,GW_2002,ZW_2006,MP_2012}. Rolling window uses a fixed sample length for fitting the model and, then, the~following step is forecasted. As~in this paper the window size is set to \(650\)~and the forecast horizon to \(1\), the~proposed and benchmark models are fitted using the last \(650\) S\&P trading days and, then, the~next day volatility is forecasted. This process is repeated until the whole period under analysis is forecasted. The~periods used as training and testing set will be defined at the end of this~subsection.

The input variables of the models proposed are the daily logarithmic returns (\(r_{t-i}\)) and the standard deviation of the last five daily logarithmic returns:
\begin{align}
\sigma_{t-1}= \sqrt{\frac{\sum_{i=1}^{n}{\big(r_{t-i}-E[r]}\big)^2}{n-1}}
\end{align}

{As} %MDPI: we added indent, please confirm.
Multi-Transformer, Transformer and LSTM layers are able to manage time series, a~lag of the last 10 observations of the previous variables are taken into consideration for fitting these layers. Thus, the~input variables are:
\begin{align}
X_1= (\sigma_{t-1}, \sigma_{t-2}, \dots, \sigma_{t-10}) \\
X_2= (r_{t-1}, r_{t-2}, \dots, r_{t-10})
\end{align}

{In} accordance with other studies such as~\cite{Roh_2006} or~\cite{RAN_2019}, the~realized volatility is used as response variable for the models based on ANNs;
\begin{align}
Y=\hat{\sigma}_{i,t}= \sqrt{\frac{\sum_{n=0}^{i-1}{(r_{t+n}-E[r_f]})^2}{i-1}}
\end{align}
where \(E[r_f]=\sum_{n=0}^{i-1}{r_{t+n}}/i\) and \(i=5\). As~shown in the previous formula, the~realized volatility can be defined as the standard deviation of future logarithmic~returns.

The dataset for fitting and evaluating the volatility forecasting models contains market data of S\&P from 1 January 2008 to 31 December 2020. The~optimum configuration of the models is obtained by applying the rolling window approach and selecting the configuration which minimizes the error (RMSE) in the period going from 1 January 2008 to 31 December 2015. The~optimum configuration in combination with the rolling window methodology is applied in order to forecast the volatility contained in the testing set (from 1 January 2016 to 31 December 2020). The~empirical results presented in Section~\ref{VolII_Comparison} are based on the forecasts of the testing~set.

\subsection{Models~Validation}
\label{VolII_Validation}
This subsection presents the measures selected for validating and comparing the performance of the benchmark models with the algorithms proposed in this~paper.

The mean absolute value (\(MAE\)) and the root mean squared error (\(RMSE\)) have been selected for validating the forecasting power of the different stock volatility models:
\begin{equation}
MAE=\sum_{t=1}^N \frac{\mid\sigma_{i,t} - \hat{\sigma}_{i,t}\mid}{N}  \qquad  {\rm/} \quad RMSE=\sum_{t=1}^N \frac{(\sigma_{i,t} - \hat{\sigma}_{i,t})^2}{N}
\end{equation}
where \(N\) is the total number of~observations.

The validation carried out by this study is not only interested on the accuracy, but~also on the appropriateness of the risk measures generated by the different stock volatility forecasting models. In~accordance with Solvency II Directive, \(99.5\%\) VaR has been selected as risk measure. Although~Solvency II has the aim of obtaining the yearly VaR, the~calculations carried out in this paper will be based on a daily VaR in order to have more data points and, thus, more robust conclusions on the performance of the different models. The~parametric approach developed by~\cite{Kupiec_1995} is used for validating the different VaR estimations. The~aim of this test is accepting (or rejecting) the hypothesis that the number of VaR exceedances are aligned with the confidence level selected for calculating the risk measure. In~addition to the previous test, the~approach suggested by~\cite{Christoffersen_1997} is also applied in order to validate the appropriateness of~VaR.

\subsection{Benchmark~Models}
\label{VolII_Benchmark}
This subsection introduces the benchmark models used in this paper: GARCH, EGARCH, AVGARCH, GJR-GARCH, TrARCH, FIGARCH and two architectures that combine GARCH-based algorithms with ANN and LSTM, respectively. The~GARCH-based algorithms will be fitted assuming that innovations, \(\epsilon_t\), follow a Student's t-distribution. Thus, the~returns generated by these models follow a conditional t-distribution \cite{Bauwens_2012}.

The generalized autoregressive conditional heteroskedasticity (GARCH) model developed by~\cite{Bollerslev_1986} has been widely used for stock volatility forecasting purposes. GARCH(p,q) has the following expression:
\begin{equation}
\hat{\sigma}_t^2 =  \omega + \sum_{i=1}^{q}{\alpha_i r_{t-i}^2} + \sum_{i=1}^{p}{ \beta_i \sigma_{t-i}^2}  \qquad  {\rm/} \quad \hat{r}_t =  \hat{\sigma}_t \epsilon_t
\end{equation}
\textls[-5]{where \(\omega_i\), \(\alpha_i\) and \(\beta_i\) are the parameters to be estimated, \(r_{t-i}\) the previous returns and \(\sigma_{t-i}^2\) the last observed volatility. As~previously stated, innovations (\(\epsilon_t\)) follow a \mbox{Student's~t-distribution.}}

The absolute value GARCH \cite{Taylor_1986}, AVGARCH(p,q), is similar to the traditional GARCH model. In~this case, the~absolute value of previous return and volatility is taken into consideration to forecast volatility:
\begin{align}
\hat{\sigma}_t =  \omega + \sum_{i=1}^{q}{\alpha_i \mid r_{t-i} \mid} + \sum_{i=1}^{p}{ \beta_i \sigma_{t-i}}
\end{align}

{As} volatility behaves differently depending on the market tendency, models such as EGARCH, GJR-GARCH or TrGARCH  were developed. EGARCH(p,q) \cite{Nelson_1991} has the following expression for the logarithm of stocks volatility:
\begin{align}
&\log{\hat{\sigma}_t^2} =  \omega + \sum_{i=1}^{p}{ \alpha_i \log{\hat{\sigma}_{t-i}^2}} +  \sum_{i=1}^{q}{ (\beta_i e_{t-i}+\gamma_i (\mid e_{t-i} \mid-E ( \mid e_{t-i} \mid)))}
\end{align}
where \(\omega_i\), \(\alpha_i\), \(\beta_i\) and \(\gamma_i\) are the parameters to be estimated and \(e_t=r_t / \sigma_t \). The~GJR-GARCH(p,o,q) developed by~\cite{gjr_1993} has the following expression:
\begin{align}
\hat{\sigma}_t^2= \omega + \sum_{i=1}^{q}{\alpha_i r_{t-i}^2} + \sum_{i=1}^{o}{\gamma_i r_{t-i}^2 I_{[r_{t-1}<0]}} + \sum_{i=1}^{p}{\beta_i \sigma_{t-i}^2}
\end{align}

{As} with EGARCH model, \(\omega_i\), \(\alpha_i\), \(\beta_i\) and \(\gamma_i\) are the parameters to be estimated. \(I_{[r_{t-1}<0]}\) takes the value of \(1\) when the subscript condition is met. Otherwise \(I_{[r_{t-1}<0]}=0\). The~volatility of the Threshold GARCH(p,o,q) (TrGARCH) model is obtained as follows:
\begin{align}
\hat{\sigma}_t= \omega + \sum_{i=1}^{q}{ \alpha_i \mid r_{t-i} \mid } + \sum_{i=1}^{o}{ \gamma_i \mid r_{t-i} \mid I_{[r_{t-i}<0]}} + \sum_{i=1}^{p}{ \beta_i \sigma_{t-i} }
\end{align}

{As} with the previous two architectures, \(\omega_i\), \(\alpha_i\), \(\beta_i\) and \(\gamma_i\) are the model parameters. The~last GARCH-based algorithm used in this paper is the fractionally integrated GARCH (FIGARCH) model developed by~\cite{BBH_1996}. The~conditional variance dynamic is
\begin{align}
\hat{\sigma}_t= \omega + \left[ 1-\beta L - \phi L (1-L)^d \right] \epsilon^2_t + \sigma h_{t-1}
\end{align}
where \(L\) is the lag operator and \(d\) the fractional differencing~parameter.

\begin{figure}[H]
%\begin{center}
\includegraphics[width=0.74\textwidth]{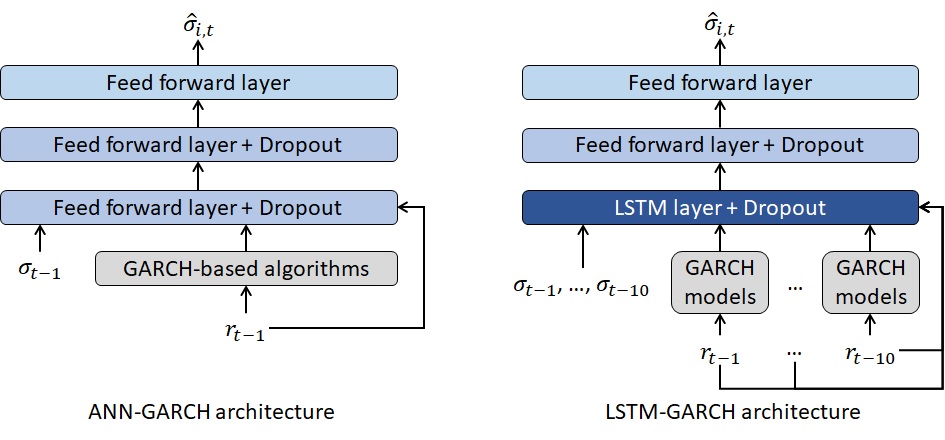}
\caption{ANN-GARCH and LSTM-GARCH architectures.}
\label{fig:Benchmark_Vol_II}
%\end{center}
\end{figure}

In addition to the previous approaches, two other hybrid models based on merging autoregressive algorithms with ANNs and LSTMs are also used as benchmark. Figure~\ref{fig:Benchmark_Vol_II} shows the architecture of ANN-GARCH and LSTM-GARCH. The~inputs of the algorithms are the~following:
\begin{itemize}
\item The last daily logarithmic return, \(r_{t-1}\), for~the ANN-GARCH and the last ten in the case of the LSTM-GARCH (as explained in Section~\ref{VolII_Data}).
\item The standard deviation of the last five daily logarithmic returns:
\begin{align}
\sigma_{t-1}= \sqrt{\frac{\sum_{i=1}^{n}{(r_{t-i}-E[r]})^2}{n-1}}
\end{align}
where \(E[r]=\sum_{i=1}^n r_{t-i} / n\) and \(n=5\). As~with the previous input variable, the~last standard deviation is considered in the ANN-GARCH, whereas the last ten are taken into consideration by the LSTM-GARCH architecture.
\end{itemize}

The GARCH-based algorithms included within the ANN-GARCH and LSTM-GARCH models are the six algorithms previously presented in this same subsection (GARCH, EGARCH, AVGARCH, GJR-GARCH, TrARCH, FIGARCH).

As explained in Section~\ref{VolII_Data}, the~true implied volatility, \(\sigma_{i,t}\), is used as response variable to train the models. This variable is the standard deviation of the future logarithmic returns:
\begin{align}
\hat{\sigma}_{i,t}= \sqrt{\frac{\sum_{n=0}^{i-1}{(r_{t+n}-E[r_f]})^2}{i-1}}
\end{align}
where \(E[r_f]=\sum_{n=0}^{i-1}{r_{t+n}}/i\). In~this paper, \(i=5\).

As it is shown in Figure~\ref{fig:Benchmark_Vol_II}, the~input of the ANN-GARCH model is processed by two feed forward layers with dropout regularization. These layers have 16 and 8 neurons, respectively. The~final output is produced by a feed forward layer with one neuron. In~the case of the LSTM-GARCH, inputs are processed by a LSTM layer with 32 units and two feed forward layers with 8 and 1 neurons, respectively, in order to produce the final~forecast.

\subsection{Transformer-Based~Models}
\label{VolII_Transformer}
\begin{figure}[H]
%\begin{center}
\includegraphics[width=0.74\textwidth]{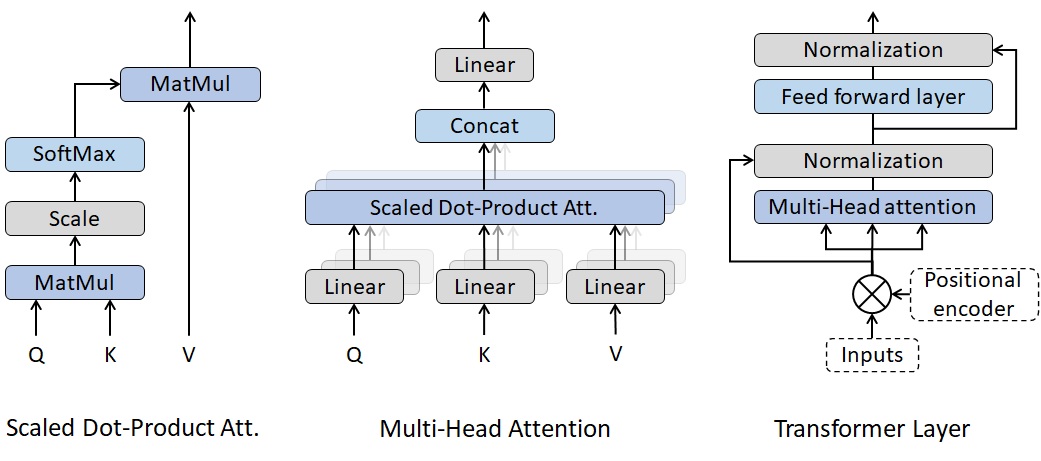}
\caption{Transformer and Multi-Head attention mechanism.}
\label{fig:Attention_Vol_II}
%\end{center}
\end{figure}

Before explaining the volatility models based on Transformer layers (see Figure~\ref{fig:Attention_Vol_II}), all the modifications applied to their architecture are presented in this subsection. As~previously stated, Transformer layers \cite{VSP_2017} were developed for NLP purposes. Thus, some modifications are needed in order to apply this layer for volatility forecasting~purposes.

In contrast to LSTM, recurrence is not present in the architecture of Transformer layers. The~two main components used by these layers in order to deal with time series are the~following:
\begin{itemize}
\item Positional encoder. As~previously stated, Transformer layers have no recurrence structure. Thus, the~information about the relative position of the observations within the time series needs to be included in the model. To~do so, a~positional encoding is added to the input data. In~the context of NLP, \citet{VSP_2017} {suggested} the following wave functions as positional encoders:
\begin{gather}
PE_{(pos,2_i)}=\sin(pos/1000^{2i/dim}) \\
PE_{(pos,2_{i+1})}=\cos(pos/1000^{2i/dim})
\end{gather}
where \(dim\) is the total number of explanatory variables (or word embedding dimension in NLP) used as input in the model, \(pos\) is the position of the observation within the time series and \(i=(1, 2, \dots, dim-1)\). This positional encoder modifies the input data depending on the lag of the time series and the embedding dimension used for the~words.

As volatility models do not use words as inputs, the~positional encoder is modified in order to avoid any variation of the inputs depending on the number of time series used as input. Thus, the~positional encoder suggested in this paper changes depending on the lag, but~it remains the same across the different explanatory variables introduced in the model. As~in the previous case, a~wave function plays the role of positional~encoder:
\begin{gather}
PE_{pos}=\cos \left(\pi \frac{pos}{N_{pos}-1} \right)= \sin \left(\frac{\pi}{2} + \pi \frac{pos}{N_{pos}-1} \right)
\end{gather}
where \(pos=(0, 1, \dots, N_{pos}-1)\) is the position of the observation within the time series and \(N_{pos}\) maximum lag.
\item Multi-Head attention. It can be considered the key component of the Transformer layers proposed by~\cite{VSP_2017}. As~shown in Figure~\ref{fig:Attention_Vol_II}, Multi-Head attention is composed of several scaled dot-product attention units running in parallel. Scaled dot-product attention is computed as follows:
\begin{gather}
Attention(Q,K,V)=softmax \left( \frac{QK^T}{\sqrt{d_k}} \right) V
\end{gather}
where \(Q\), \(K\) and \(V\) are input matrices and \(d_k\) the number of input variables taken into consideration within the dot-product attention mechanism. Multi-Head attention splits the explicative variables in different groups or `heads' in order to run the different scaled dot-product attention units in parallel. Once the different heads are calculated, the~outputs are concatenated (\textit{Concat} operator) and connected to a feed forward layer with linear activation. Thus, the~Multi-Head attention mechanism has the following expression:
\begin{gather}
MultiHead(Q,K,V)=Concat \left( head_1, \dots , head_h \right)W^O \\
head_i=Attention(QW^Q_i,KW^K_i,VW^V_i)
\end{gather}
where \(h\) is the number of heads. It is also worth mentioning that all the matrices of parameters (\(W^Q_i\), \(W^K_i\), \(W^V_i\) and \(W^O\)) are trained using feed forward layers with linear~activations.
\end{itemize}

{In} addition to the scaled dot-product and the Multi-Head attention mechanisms, \mbox{Figure~\ref{fig:Attention_Vol_II}} shows the Transformer layers used in this paper. As~suggested by~\cite{VSP_2017}, the~Multi-Head attention is followed by a normalization, a~feed forward layer with ReLU activation and, again, a~normalization layer. Transformer layers also include two residual connections~\cite{HZRS_2016}. Thanks to these connections, the~model will decide by itself if the training of some layers needs to be skipped during some phases of the fitting~process.

The modified version of Transformer layers explained in the previous paragraphs are used in the volatility models presented in Figure~\ref{fig:T_Models_Vol_II}. The~T-GARCH architecture proposed in this paper merges the six GARCH algorithms presented in Section~\ref{VolII_Benchmark} with Transformer and feed forward layers in order to forecast \(\hat{\sigma}_{i,t}\). In~addition to the previous algorithms and layers, TL-GARCH includes a LSTM with 32 units. In~this last model, the~temporal structure of the data is recognized and modelled by the LSTM layer and, thus, no positional encoder is needed in the Transformer layer. Both models have the following~characteristics:
\begin{figure}[H]
%\begin{center}
\includegraphics[width=0.70\textwidth]{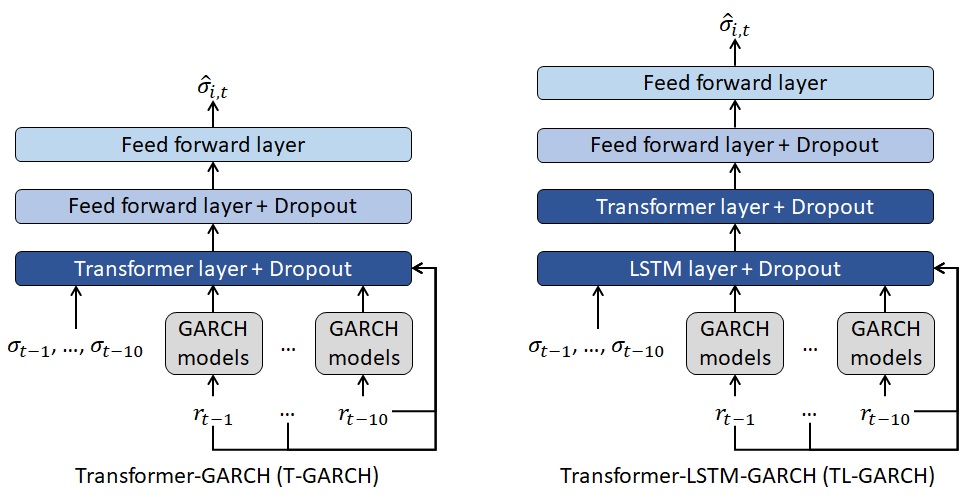}
\caption{T-GARCH and TL-GARCH volatility models.}
\label{fig:T_Models_Vol_II}
%\end{center}
\end{figure}
\vspace{-6pt}
\begin{itemize}
\item Adaptative Moment Estimator (ADAM) is the algorithm used for updating the weights of the feed forward, LSTM and Transformer layers. This algorithm takes into consideration current and previous gradients in order to implement a progressive adaptation of the initial learning rate. The~values suggested by~\cite{dk_2014} for the ADAM parameters are used in this paper and the initial learning rate is set to \(\delta=0.01\).
\item The feed forward layers with dropout present in both models have 8 neurons, while the output layer has just one.
\item The level of dropout regularization \(\theta\) \cite{SHK_2014} is optimized with the training set mentioned in Section~\ref{VolII_Data}.
\item The loss function used for weights optimization and back propagation purposes is the mean squared error.
\item Batch size is equal to 64 and the models are trained during 5000 epochs in order to obtain the final weights.
\end{itemize}

\subsection{Multi-Transformer-Based~Models}
\label{VolII_MultiTransformer}
\begin{figure}[H]
%\begin{center}
\includegraphics[width=0.70\textwidth]{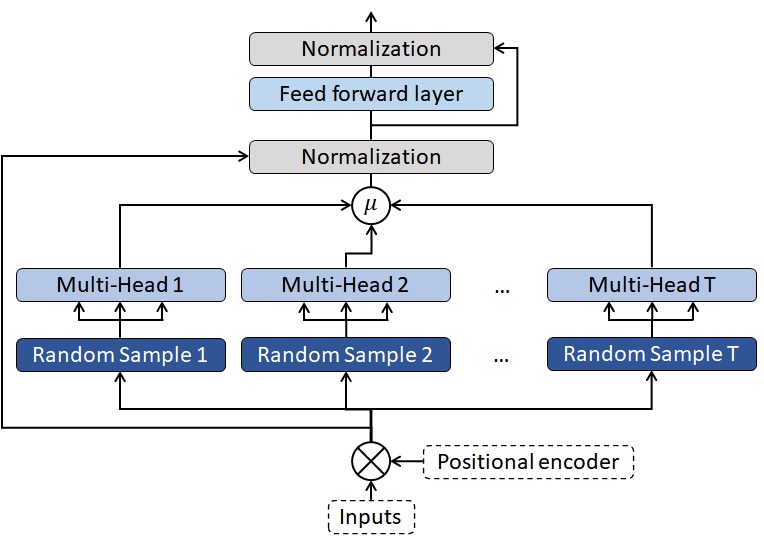}
\caption{Multi-Transformer architecture.}
\label{fig:MTL_Vol_II}
%\end{center}
\end{figure}

This subsection presents the Multi-Transformer layers and the volatility models based on them. The~Multi-Transformer architecture proposed in this paper is a variant of the Transformer layers proposed by~\cite{VSP_2017}. The~main differences between both architectures are the~following:

\begin{itemize}
\item As shown in Figure~\ref{fig:MTL_Vol_II}, Multi-Transformer layers generate \(T\) different random samples of the input data. In~the volatility models proposed in this paper, \(90\%\) of the observations of the database are randomly selected in order to compute the different~samples.
\item Multi-Transformer architecture is composed of \(T\) Multi-Head attention units (in this paper \(T=5\)), one per each random sample of the input data. Then, the~average of the different units is computed in order to obtain the final attention matrix. Thus, the~Average Multi-Head (AMH) mechanism present in Multi-Transformer can be defined as~follows:
\begin{gather}
AMH(Q,K,V)=\frac{\sum^T_{t=1}  Concat \left( head_{1,t}, \dots , head_{h,t} \right)W^O_t}{T} \\
head_{i,t}=Attention(Q_t W^Q_{i,t},K_t W^K_{i,t},V_t W^V_{i,t})
\end{gather}
\end{itemize}

{As} with the Transformer architecture applied in this paper, the~positional encoder used is \(PE_{pos}\) instead of \(PE_{(pos,2_i)}\) and \(PE_{(pos,2_{i+1})}\).
The aim of the Multi-Transformer layers introduced in the paper is to improve the stability and accuracy by applying bagging \cite{Breiman_1996} to the attention mechanism. This technique is usually applied to algorithms such as linear regression, neural networks or decision trees. Instead of applying the procedure on all the data that are input into the model, the~proposed methodology uses bagging only to the attention mechanism of the layer~architecture.

The computational power required by bagging is one of the main limitations of this technique. As~Multi-Transformer applies bagging to the attention mechanisms, their weights are trained several times in each epoch. Nevertheless, bagging is not applied to the rest of the layer weights and, thus, this offsets partially the previous limitation. It is also worth mentioning that bagging preserves the bias and this may result in~underfitting.

On the other hand, this technique should bring two main advantages to the Multi-Transformer layer. First, bagging reduces significantly the error variance. Second, the~aggregation of learners using this technique leads to a higher accuracy and reduces the risk of~overfitting.

\begin{figure}[H]
%\begin{center}
\includegraphics[width=0.70\textwidth]{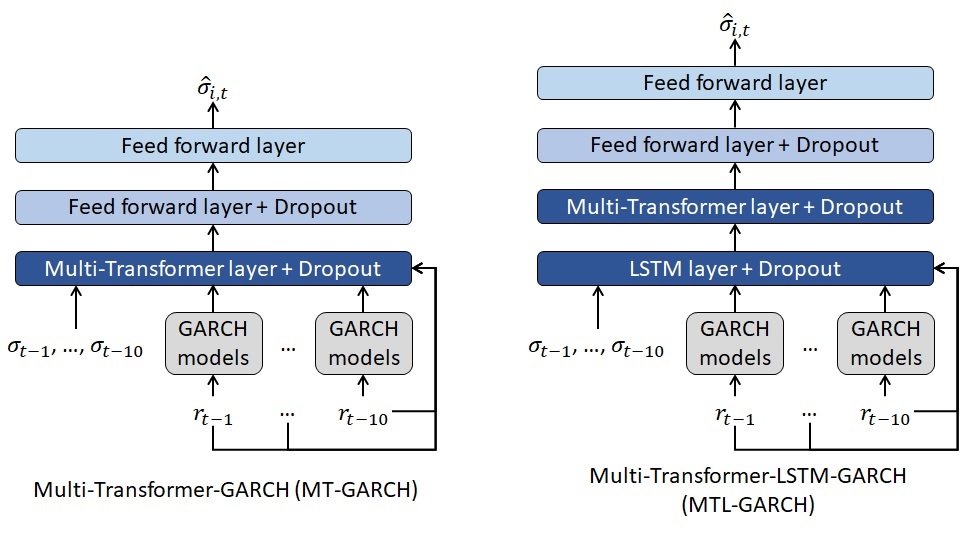}
\caption{MT-GARCH and MTL-GARCH volatility models.}
\label{fig:MT_Models_Vol_II}
%\end{center}
\end{figure}

The structure of the volatility models based on Multi-Transformer layers (Figure \ref{fig:MT_Models_Vol_II}) is similar to the architectures presented in Section~\ref{VolII_Transformer}. The~MT-GARCH merges Multi-Transformer and feed forward layers with the six GARCH models presented in \mbox{Section~\ref{VolII_Benchmark}}. In~addition to the previous algorithms and layers, MTL-GARCH adds a LSTM with 32~units. The~rest of the characteristics such as the optimizer, the~number of neurons of the feed forward layers or the level of dropout regularization are the same than those presented in the previous section for T-GARCH and~TL-GARCH.

The risk measures of ANN-GARCH, LSTM-GARCH and all the models introduced by this paper (Sections \ref{VolII_Transformer} and \ref{VolII_MultiTransformer}) are calculated assuming that daily log-returns follow a non-standardize Student's t-distribution with standard deviation equal to the forecasts made by the volatility models. It is worth mentioning that Student's t-distribution generates more appropriate risk measures than normal distribution due to the shape of its tail \cite{JT_2014,McNeil_2015}. In~addition, this assumption is in line with the GARCH-based models used as benchmark and the inputs of the hybrid models presented in this~paper.

%Seccion de resultados
\section{Results}
\label{VolII_resul}
In this section, the~forecasts and the risk measures of the volatility models presented in previous sections are compared with the ones obtained from the benchmark models. In~addition, the~following subsection shows the optimum hyperparameters of the benchmark and proposed hybrid volatility~models.
\subsection{Fitting of Models Based on Neural~Networks}
\label{VolII_Fitting}
As explained in Section~\ref{VolII_Data}, rolling window approach (\cite{S_1998,GW_2002,ZW_2006,MP_2012} among others) is applied for fitting the algorithms. The~training set used for optimizing the level of dropout regularization contains S\&P returns and observed volatilities from 1 January 2008 to 31 December 2015. Table~\ref{VolII_Hyperparameter} presents the error by model and level of \(\theta\).

\begin{specialtable}[H]
%\begin{center}
\caption{RMSE by level of \(\theta\).}
\label{VolII_Hyperparameter}

\setlength{\cellWidtha}{\columnwidth/5-2\tabcolsep+0.0in}
\setlength{\cellWidthb}{\columnwidth/5-2\tabcolsep+0.0in}
\setlength{\cellWidthc}{\columnwidth/5-2\tabcolsep+0.0in}
\setlength{\cellWidthd}{\columnwidth/5-2\tabcolsep+0.0in}
\setlength{\cellWidthe}{\columnwidth/5-2\tabcolsep+0.0in}
\scalebox{1}[1]{\begin{tabularx}{\columnwidth}{>{\PreserveBackslash\centering}m{\cellWidtha}>{\PreserveBackslash\centering}m{\cellWidthb}>{\PreserveBackslash\centering}m{\cellWidthc}>{\PreserveBackslash\centering}m{\cellWidthd}>{\PreserveBackslash\centering}m{\cellWidthe}}

\toprule
\textbf{Model}      & \boldmath{\(\theta=0\)}    & \boldmath{\(\theta=0.05\)}  & \boldmath{\(\theta=0.10\)} & \boldmath{\(\theta=0.15\)}  \\
\midrule
ANN-GARCH  & 0.0351          & 0.0092           & 0.0085          & 0.0082           \\
LSTM-GARCH & 0.0065          & 0.0057           & 0.0056          & 0.0054           \\
T-GARCH    & 0.0089          & 0.0076           & 0.0072          & 0.0074           \\
TL-GARCH   & 0.0050          & 0.0045           & 0.0044          & 0.0045           \\
MT-GARCH   & 0.0068          & 0.0062           & 0.0064          & 0.0064           \\
MTL-GARCH  & 0.0047          & 0.0045           & 0.0042          & 0.0044           \\
\bottomrule
\end{tabularx}}
{\footnotesize \noindent {\emph{Source}: own elaboration. }}
%\end{center}
\end{specialtable}

\vspace{-6pt}

The results of the optimization process reveals that \(\theta=0\) generates higher error rates than the rest of the possible values regardless of the model. This means that models based on architectures such as Transformer, LSTM or feed forward layers need an appropriate level of regularization in order to avoid overfitting. According to the results, this is especially relevant for ANN-GARCH, where the error strongly depends on the level of regularization. The~dropout level that minimizes the error of each model is~selected.

\subsection{Comparison against Benchmark~Models}
\label{VolII_Comparison}
Once the optimum dropout level of each of the proposed volatility forecasting models based on Transformer and Multi-Transformer is selected, their performance is compared with the benchmark models (traditional GARCH processes, ANN-GARCH and LSTM-GARCH) presented in Section~\ref{VolII_Benchmark}.

Tables~\ref{RMSE_VolII} and \ref{MAE_VolII} present the validation error (RMSE and MAE) by year and model. The~column `Total' shows the error of the whole test period (from \mbox{1 January 2016} to\linebreak \textls[-5]{\mbox{31 December 2020}). The~main conclusions drawn from the these tables are the following:}

\begin{itemize}
\item Traditional GARCH processes are outperformed by models based on merging artificial neural network architectures such as feed forward, LSTM or Transformer layers with the outcomes of autoregressive algorithms (also named hybrid models).
\item The comparison between ANN-GARCH and the rest of the volatility forecasting models based on artificial neural networks (LSTM-GARCH, T-GARCH, TL-GARCH, MT-GARCH and MTL-GARCH) reveals that feed forward layers lead to less accurate forecasts than other architectures. Multi-Transformer, Transformer and LSTM were specially created to forecast time series and, thus, the~volatility models based on these layers are more accurate than ANN-GARCH.
\item Merging Multi-Transformer and Transformer layers with LSTMs leads to more accurate predictions than traditional LSTM-based architectures. Indeed, TL-GARCH achieves better results than LSTM-GARCH, even though the number of weights of TL-GARCH is significantly lower. Thus, the~novel Transformer and Multi-Transformer layers introduced for NLPs purposes can be adapted as described in \mbox{Sections~\ref{VolII_Transformer} and \ref{VolII_MultiTransformer}} in order to generate more accurate volatility forecasting models. It is also worth mentioning that Multi-Transformer layers, which were also introduced in this paper, lead to more accurate forecasts thanks to their ability to average several attention mechanisms. In~fact, the~model that achieves the lower MAE and RMSE is a mixture of Multi-Transformer and LSTM layers (MTL-GARCH).
\end{itemize}

\begin{specialtable}[H]
%\begin{center}
\caption{RMSE by volatility model and~year.}
\label{RMSE_VolII}
\setlength{\cellWidtha}{\columnwidth/7-2\tabcolsep+0.6in}
\setlength{\cellWidthb}{\columnwidth/7-2\tabcolsep-0.1in}
\setlength{\cellWidthc}{\columnwidth/7-2\tabcolsep-0.1in}
\setlength{\cellWidthd}{\columnwidth/7-2\tabcolsep-0.1in}
\setlength{\cellWidthe}{\columnwidth/7-2\tabcolsep-0.1in}
\setlength{\cellWidthf}{\columnwidth/7-2\tabcolsep-0.1in}
\setlength{\cellWidthg}{\columnwidth/7-2\tabcolsep-0.1in}
\scalebox{1}[1]{\begin{tabularx}{\columnwidth}{>{\PreserveBackslash\centering}m{\cellWidtha}>{\PreserveBackslash\centering}m{\cellWidthb}>{\PreserveBackslash\centering}m{\cellWidthc}>{\PreserveBackslash\centering}m{\cellWidthd}>{\PreserveBackslash\centering}m{\cellWidthe}>{\PreserveBackslash\centering}m{\cellWidthf}>{\PreserveBackslash\centering}m{\cellWidthg}}

\toprule
\textbf{Model}           & \textbf{2016}    & \textbf{2017}    & \textbf{2018}    & \textbf{2019}    & \textbf{2020}    & \textbf{Total}   \\
\midrule
{GARCH(1,1)} %MDPI: please confirm if space should be added before "('', and after ",''.  Authors: GARCH(1,1) is normally written with no spaces. This applies for all the other autoregressive models.
& 0.0058  & 0.0026  & 0.0095  & 0.0073  & 0.1026  & 0.0464  \\
AVGARCH(1,1)    & 0.0053  & 0.0027  & 0.0076  & 0.0056  & 0.0847  & 0.0383  \\
EGARCH(1,1)     & 0.0056  & 0.0028  & 0.0093  & 0.0078  & 0.0880  & 0.0399  \\
GJR-GARCH(1,1,1)& 0.0090  & 0.0028  & 0.0126  & 0.0068  & 0.1248  & 0.0565  \\
TrGARCH(1,1,1)  & 0.0074  & 0.0027  & 0.0115  & 0.0058  & 0.1153  & 0.0521  \\
FIGARCH(1,1)    & 0.0062  & 0.0029  & 0.0095  & 0.0066  & 0.1011  & 0.0457  \\
ANN-GARCH       & 0.0042  & 0.0023  & 0.0060  & 0.0044  & 0.0171  & 0.0086  \\
LSTM-GARCH      & 0.0032  & 0.0021  & 0.0043  & 0.0030  & 0.0101  & 0.0054  \\
T-GARCH         & 0.0048  & 0.0029  & 0.0058  & 0.0044  & 0.0117  & 0.0067  \\
TL-GARCH        & 0.0030  & 0.0019  & 0.0033  & 0.0026  & 0.0070  & 0.0040  \\
MT-GARCH        & 0.0036  & 0.0021  & 0.0046  & 0.0033  & 0.0096  & 0.0054  \\
MTL-GARCH       & 0.0030  & 0.0016  & 0.0033  & 0.0026  & 0.0066  & 0.0038  \\
\bottomrule
\end{tabularx}}
{\footnotesize \noindent {\emph{Source}: own elaboration. }}
\end{specialtable}
\unskip

\begin{specialtable}[H]
%\begin{center}
\caption{MAE by volatility model and~year.}
\label{MAE_VolII}
\setlength{\cellWidtha}{\columnwidth/7-2\tabcolsep+0.6in}
\setlength{\cellWidthb}{\columnwidth/7-2\tabcolsep-0.1in}
\setlength{\cellWidthc}{\columnwidth/7-2\tabcolsep-0.1in}
\setlength{\cellWidthd}{\columnwidth/7-2\tabcolsep-0.1in}
\setlength{\cellWidthe}{\columnwidth/7-2\tabcolsep-0.1in}
\setlength{\cellWidthf}{\columnwidth/7-2\tabcolsep-0.1in}
\setlength{\cellWidthg}{\columnwidth/7-2\tabcolsep-0.1in}
\scalebox{1}[1]{\begin{tabularx}{\columnwidth}{>{\PreserveBackslash\centering}m{\cellWidtha}>{\PreserveBackslash\centering}m{\cellWidthb}>{\PreserveBackslash\centering}m{\cellWidthc}>{\PreserveBackslash\centering}m{\cellWidthd}>{\PreserveBackslash\centering}m{\cellWidthe}>{\PreserveBackslash\centering}m{\cellWidthf}>{\PreserveBackslash\centering}m{\cellWidthg}}
\toprule
\textbf{Model}           & \textbf{2016}    & \textbf{2017}    & \textbf{2018}    & \textbf{2019}    & \textbf{2020}    & \textbf{Total}   \\
\midrule
{GARCH(1,1)}      & 0.0037  & 0.0019  & 0.0058  & 0.0044  & 0.0363  & 0.0105  \\
AVGARCH(1,1)    & 0.0034  & 0.0019  & 0.0049  & 0.0037  & 0.0296  & 0.0087  \\
EGARCH(1,1)     & 0.0035  & 0.0020  & 0.0060  & 0.0048  & 0.0333  & 0.0100  \\
GJR-GARCH(1,1,1)& 0.0048  & 0.0020  & 0.0074  & 0.0042  & 0.0404  & 0.0118  \\
TrGARCH(1,1,1)  & 0.0042  & 0.0020  & 0.0069  & 0.0038  & 0.0365  & 0.0107  \\
FIGARCH(1,1)    & 0.0038  & 0.0021  & 0.0055  & 0.0041  & 0.0361  & 0.0104  \\
ANN-GARCH       & 0.0029  & 0.0019  & 0.0038  & 0.0029  & 0.0095  & 0.0042  \\
LSTM-GARCH      & 0.0022  & 0.0015  & 0.0027  & 0.0021  & 0.0060  & 0.0029  \\
T-GARCH         & 0.0035  & 0.0021  & 0.0041  & 0.0031  & 0.0070  & 0.0040  \\
TL-GARCH        & 0.0020  & 0.0014  & 0.0021  & 0.0018  & 0.0044  & 0.0023  \\
MT-GARCH        & 0.0024  & 0.0016  & 0.0031  & 0.0023  & 0.0057  & 0.0030  \\
MTL-GARCH       & 0.0019  & 0.0012  & 0.0021  & 0.0018  & 0.0041  & 0.0022  \\
\bottomrule
\end{tabularx}}
{\footnotesize \noindent {\emph{Source}: own elaboration. }}
\end{specialtable}

\vspace{-6pt}

\begin{figure}[H]
%\begin{center}
\includegraphics[width=0.70\textwidth]{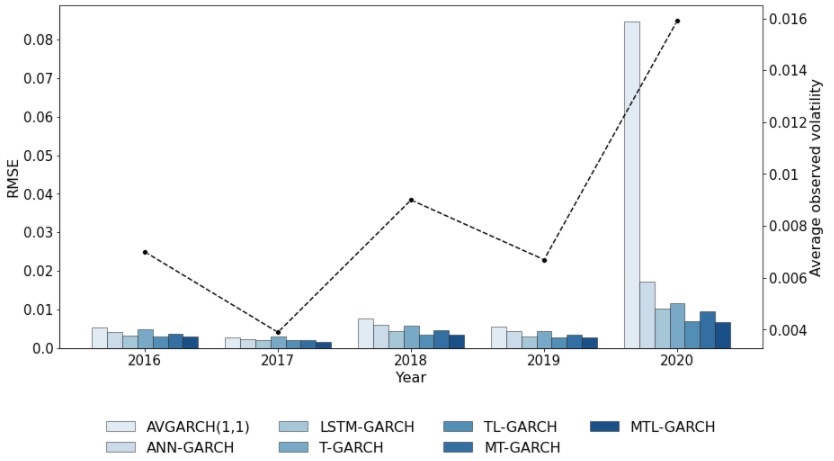}
\caption{Observed volatility and RMSE by year.}
\label{fig:Error_Vol_II}
%\end{center}
\end{figure}

To enhance the analysis of the results shown in Tables~\ref{RMSE_VolII} and \ref{MAE_VolII}, Figure~\ref{fig:Error_Vol_II} collects the RMSE and the observed volatility by year. Notice that only the most accurate GARCH-based model is shown in order to improve the visualization of the graph. The~black dashed line shows that the observed volatility of 2020 was significantly higher than the rest of the years due to the turmoil caused by COVID-19 outbreak. As~expected, the~error of every model is also higher in 2020 because the market volatility was more unpredictable than the rest of the years. Nevertheless, it has to be mentioned that the 2020 forecasts of traditional autoregressive algorithms are significantly less accurate than hybrid models based on architectures such as LSTM, Transformer or Multi-Transformer~layers.

Although the observed volatility is lower in years before 2020, autoregressive models are also outperformed by hybrid models. Nevertheless, the~difference between both sets of models is remarkably~lower.

The \emph{p}-values of the Kupiec and Christoffersen tests by volatility model and year are shown in Tables~\ref{Kupiec_VolII} and \ref{Chris_VolII}, respectively. In~contrast to the approach suggested by Kupiec, Christoffersen test is not only focused on the total number of exceedances, but~it also takes into consideration the number of consecutive VaR exceedances. As~stated in Section~\ref{VolII_Validation}, the~risk measure and confidence level (\(99.5\%\) VaR) selected are in line with Solvency II Directive. This regulation sets the principles for calculating the capital requirements and assessing the risk profile of the insurance companies based in the European Union. This law covers not only the underwriting risks but also financial risks such as the potential losses due to variations on the interest rate curves or the equity~prices.

The column ‘Total’ of Tables~\ref{Kupiec_VolII} and \ref{Chris_VolII} reveal that only TL-GARCH, MT-GARCH and MTL-GARCH produce appropriate risk measures (\emph{p}-value higher than 0.05 in both tests) for the period 2016--2020. The~rest of the models fail both tests and, thus, their risk measures can not be considered to be appropriate for that~period.

As with any other statistical test, the~higher the number of data points the more relevant are the outcomes obtained from the test. That is the reason why the previous paragraph focuses on the ‘Total’ column and not on the specific results obtained by year. The~results by year show that most of the models fail the test in 2020 due to the high level of volatility produced by COVID-19~pandemic.

According to these results, the~stock volatility models introduced in this paper (T-GARCH, TL-GARCH, MT-GARCH and MTL-GARCH) produce more accurate estimations and appropriate risk measures in most of the cases. Regarding the models accuracy, it is specially remarkable the difference observed in 2020, where COVID-19 caused a significant turmoil in the stock market. Concerning the appropriateness of equity risk measures, three out of four models based on Transformer and Multi-Transformer pass Kupiec and Christofferesen test for the period 2016--2020, while all the benchmark models fail at least one of them. Notice that the proposed models are compared with other approaches belonging to its own family (ANN-GARCH and LSTM-GARCH) and autoregressive models belonging to the GARCH~family.
\begin{specialtable}[H]
%\begin{center}
\caption{Kupiec test (\emph{p}-values) by volatility model and~year.}
\label{Kupiec_VolII}
\setlength{\cellWidtha}{\columnwidth/7-2\tabcolsep+0.6in}
\setlength{\cellWidthb}{\columnwidth/7-2\tabcolsep-0.1in}
\setlength{\cellWidthc}{\columnwidth/7-2\tabcolsep-0.1in}
\setlength{\cellWidthd}{\columnwidth/7-2\tabcolsep-0.1in}
\setlength{\cellWidthe}{\columnwidth/7-2\tabcolsep-0.1in}
\setlength{\cellWidthf}{\columnwidth/7-2\tabcolsep-0.1in}
\setlength{\cellWidthg}{\columnwidth/7-2\tabcolsep-0.1in}
\scalebox{1}[1]{\begin{tabularx}{\columnwidth}{>{\PreserveBackslash\centering}m{\cellWidtha}>{\PreserveBackslash\centering}m{\cellWidthb}>{\PreserveBackslash\centering}m{\cellWidthc}>{\PreserveBackslash\centering}m{\cellWidthd}>{\PreserveBackslash\centering}m{\cellWidthe}>{\PreserveBackslash\centering}m{\cellWidthf}>{\PreserveBackslash\centering}m{\cellWidthg}}
\toprule
\textbf{Model}           & \textbf{2016}    & \textbf{2017}    & \textbf{2018}    & \textbf{2019}    & \textbf{2020}    & \textbf{Total}   \\
\midrule
{GARCH(1,1)}      & 0.543 & 0.540 & 0.051 & 0.543 & 0.052 & 0.008     \\
AVGARCH(1,1)    & 0.543 & 0.540 & 0.051 & 0.543 & 0.052 & 0.008     \\
EGARCH(1,1)     & 0.543 & 0.540 & 0.051 & 0.543 & 0.052 & 0.008     \\
GJR-GARCH(1,1,1)& 0.543 & 0.540 & 0.011 & 0.543 & 0.190 & 0.008     \\
TrGARCH(1,1,1)  & 0.543 & 0.540 & 0.051 & 0.810 & 0.190 & 0.042     \\
FIGARCH(1,1)    & 0.543 & 0.540 & 0.051 & 0.543 & 0.052 & 0.008     \\
ANN-GARCH       & 0.543 & 0.540 & 0.001 & 0.002 & 0.012 & 0.001     \\
LSTM-GARCH      & 0.810 & 0.186 & 0.540 & 0.188 & 0.190 & 0.042     \\
T-GARCH         & 0.188 & 0.540 & 0.002 & 0.543 & 0.052 & 0.001     \\
TL-GARCH        & 0.543 & 0.540 & 0.813 & 0.810 & 0.810 & 0.782     \\
MT-GARCH        & 0.112 & 0.540 & 0.540 & 0.188 & 0.052 & 0.089     \\
MTL-GARCH       & 0.543 & 0.113 & 0.113 & 0.810 & 0.190 & 0.910     \\
\bottomrule
\end{tabularx}}
{\footnotesize \noindent {\emph{Source}: own elaboration. }}
\end{specialtable}
\unskip

\begin{specialtable}[H]
%\begin{center}
\caption{Christoffersen test (\emph{p}-values) by volatility model and~year.}
\label{Chris_VolII}
\setlength{\cellWidtha}{\columnwidth/7-2\tabcolsep+0.6in}
\setlength{\cellWidthb}{\columnwidth/7-2\tabcolsep-0.1in}
\setlength{\cellWidthc}{\columnwidth/7-2\tabcolsep-0.1in}
\setlength{\cellWidthd}{\columnwidth/7-2\tabcolsep-0.1in}
\setlength{\cellWidthe}{\columnwidth/7-2\tabcolsep-0.1in}
\setlength{\cellWidthf}{\columnwidth/7-2\tabcolsep-0.1in}
\setlength{\cellWidthg}{\columnwidth/7-2\tabcolsep-0.1in}
\scalebox{1}[1]{\begin{tabularx}{\columnwidth}{>{\PreserveBackslash\centering}m{\cellWidtha}>{\PreserveBackslash\centering}m{\cellWidthb}>{\PreserveBackslash\centering}m{\cellWidthc}>{\PreserveBackslash\centering}m{\cellWidthd}>{\PreserveBackslash\centering}m{\cellWidthe}>{\PreserveBackslash\centering}m{\cellWidthf}>{\PreserveBackslash\centering}m{\cellWidthg}}
\toprule
\textbf{Model}           & \textbf{2016}    & \textbf{2017}    & \textbf{2018}    & \textbf{2019}    & \textbf{2020}    & \textbf{Total}   \\
\midrule
{GARCH(1,1)}      & 0.522 & 0.520 & 0.004 & 0.523 & 0.048 & 0.002     \\
AVGARCH(1,1)    & 0.522 & 0.520 & 0.004 & 0.523 & 0.048 & 0.002     \\
EGARCH(1,1)     & 0.522 & 0.520 & 0.004 & 0.523 & 0.048 & 0.002     \\
GJR-GARCH(1,1,1)& 0.522 & 0.520 & 0.002 & 0.523 & 0.179 & 0.002     \\
TrGARCH(1,1,1)  & 0.522 & 0.520 & 0.004 & 0.800 & 0.179 & 0.009     \\
FIGARCH(1,1)    & 0.522 & 0.520 & 0.004 & 0.523 & 0.048 & 0.002     \\
ANN-GARCH       & 0.522 & 0.520 & 0.001 & 0.002 & 0.002 & 0.001     \\
LSTM-GARCH      & 0.800 & 0.180 & 0.520 & 0.177 & 0.179 & 0.037     \\
T-GARCH         & 0.176 & 0.520 & 0.001 & 0.523 & 0.048 & 0.001     \\
TL-GARCH        & 0.522 & 0.520 & 0.803 & 0.800 & 0.797 & 0.693     \\
MT-GARCH        & 0.113 & 0.520 & 0.520 & 0.177 & 0.048 & 0.079     \\
MTL-GARCH       & 0.522 & 0.113 & 0.113 & 0.800 & 0.179 & 0.790     \\
\bottomrule
\end{tabularx}}
{\footnotesize \noindent {\emph{Source}: own elaboration. }}
\end{specialtable}

\vspace{-12pt}

%Sección de debate
\section{Discussion}
\label{VolII_Conclusion}
This paper introduced a set of volatility forecasting models based on Transformer and Multi-Transformer layers. As~Transformer layers were developed for NLP purposes~\cite{VSP_2017}, their architecture is adapted in order to generate stock volatility forecasting models. Multi-Transformer layers, which are introduced by this paper, have the aim of improving the stability and accuracy of Transformer layers by applying bagging to the attention mechanism. The~predictive power and risk measures generated by the proposed volatility forecasting models (T-GARCH, TL-GARCH, MT-GARCH and MTL-GARCH) are compared with traditional GARCH processes and other hybrid models based on LSTM and feed forward~layers.

Three main outcomes were drawn from the empirical results. First, hybrid models based on LSTM, Transformer or Multi-Transformer layers outperform traditional autoregressive algorithms and hybrid models based on feed forward layers. The~validation error by year shows that this difference is more relevant in 2020, when the volatility of S\&P500 was significantly higher than in the previous years due to COVID-19 pandemic. Volatility forecasting models are mainly used for pricing derivatives and assessing the risk profile of financial institutions. As~the more relevant shocks on the solvency position of financial institutions and derivatives prices are observed in high volatility regimes, the~accurateness of these models is particularly important in years such as~2020.

The higher performance of hybrid models have also been demonstrated by~\cite{Roh_2006,hsz_2012,kfm_2014,m_2014,lu_2016,kw_2018,bk_2018}. These papers merged traditional GARCH models with feed forward layers to predict stock market volatility. This type of models have shown also a superior performance in other financial fields such as oil market volatility \cite{km_2016,Verma_2021} and metals price volatility \cite{km_2015,kh_2017}. Notice that this paper does not only present a comparison with traditional autoregressive models, but~it also shows that Transformer and Multi-Transformer can lead to more accurate volatility estimations than other hybrid~models.

Second, Multi-Transformer layers lead to more accurate volatility forecasting models than Transformer layers. As~expected, applying bagging to the attention mechanism has a positive impact on the performance of the models presented in this paper. It is also remarkable that empirical results demonstrate that merging LSTM with Transformer or Multi-Transformer layers has also a positive impact on the models performance. On~one hand, the~volatility forecasting model based on Multi-Transformer and LSTM (named MTL-GARCH) achieves the best results in the period 2016--2020. On~the other hand, the~merging of Transfomer with LSTM (TL-GARCH) leads to a lower error rate than the hybrid model based only on LSTM layers (LSTM-GARCH) even though the number of weights of the first model is significantly lower. Thus, the~use of Transfomer layers can lead to simpler and more accurate volatility forecasting models. Notice that Transformer layers are already considered the state of art thanks to BERT \cite{BERT_2018} and GPT-3 \cite{GPT3_2020}. These models have been successfully used for sentence prediction, conversational response generation, sentiment classification, coding and writing fiction, among~others.

Third, the~results of Kupiec and Christoffersen tests revealed that only the risk estimations made by MTL-GARCH, TL-GARCH and MT-GARCH can be considered as appropriate for the period 2016--2020, whereas traditional autoregressive algorithms and hybrid models based on feed forward and LSTM layers failed, at~least, one of the tests. As~previously stated, volatility does not play only a key role in risk management but also in derivative valuation models. Thus, using a volatility model that generates appropriate risk measures can lead to more accurate derivatives~valuation.

%Sección de conclusiones
\section{Conclusions}
\label{VolII_Conclusion_II}

Transformer layers are the state of the art in natural language processing. Indeed, the~performance of this layer have overcome the performance of any other previous model in this field \cite{GPT3_2020}. As~Transformer layers were specially created for natural language processing, they need to be modified in order to be used for other purposes. Probably, this is one of the main reasons why this layer have not been already extended to other fields. This paper provides the modifications needed to apply this layer for stock volatility forecasting purposes. The~results shown in this paper demonstrates that Transformer layers can overcome also the performance of the main stock volatility~models.

Following the intuition of bagging \cite{Breiman_1996}, this paper introduces Multi-Transformer layers. This novel architecture has the aim of improving the stability and accuracy of the attention mechanism, which is the core of Transformer layers. According to the results, it can be concluded that this procedure improves the accuracy of stock volatility models based on Transformer~layers.

Leaving aside the comparisons between Transformer and Multi-Transformer layers, the~hybrid models based on them have overcome the performance of autoregressive algorithms and other models based on feed forward layers and LSTMs. The~architecture of these hybrid models (T-GARCH, TL-GARCH, MT-GARCH and MTL-GARCH) based on Transformer and Multi-Transformer layers is also provided in this~paper.

According to the results, it is also worth noticing that the risk estimations based on the previous models are specially appropriate. The~VaR of most of these models can be considered accurate even in years such as 2020, when the COVID-19 pandemic caused a remarkable turmoil in the stock~market.

Consequently, the~empirical results obtained with the hybrid models based on Transfomer and Multi-Transformer layers suggest that further investigation should be conducted about the possible application of them for derivative valuation purposes. Notice that volatility plays a key role in the financial derivatives valuation. In~addition, the~models can be extended by merging Transformer or Multi-Transformer layers with other algorithms (such as gradient boosting with trees or random forest) or modifying some key assumptions of the attention~mechanism.

\vspace{6pt}

%%%%%%%%%%%%%%%%%%%%%%%%%%%%%%%%%%%%%%%%%%
\authorcontributions{Conceptualization, E.R.-P.; methodology, E.R.-P., P.J.A.-G. and J.J.N.-V.; software, E.R.-P.; validation, P.J.A.-G. and J.J.N.-V.; formal analysis, E.R.-P.; investigation, E.R.-P., P.J.A.-G. and J.J.N.-V.; writing---both original draft preparation, review and editing, E.R.-P., P.J.A.-G. and J.J.N.-V.; supervision, P.J.A.-G. and J.J.N.-V.; project administration, P.J.A.-G. and J.J.N.-V.; funding acquisition, P.J.A.-G. and~J.J.N.-V. All authors have read and agreed to the published version of the~manuscript.}

\funding{The APC was funded by Economics Department of Universidad de Alcal\'a.}

%\institutionalreview{{ }}%In this section, please add the Institutional Review Board Statement and approval number for studies involving humans or animals. Please note that the Editorial Office might ask you for further information. Please add ``The study was conducted according to the guidelines of the Declaration of Helsinki, and approved by the Institutional Review Board (or Ethics Committee) of NAME OF INSTITUTE (protocol code XXX and date of approval).'' OR ``Ethical review and approval were waived for this study, due to REASON (please provide a detailed justification).'' OR ``Not applicable'' for studies not involving humans or animals. You might also choose to exclude this statement if the study did not involve humans or animals.}
%
%\informedconsent{{ }}%Any research article describing a study involving humans should contain this statement. Please add ``Informed consent was obtained from all subjects involved in the study.'' OR ``Patient consent was waived due to REASON (please provide a detailed justification).'' OR ``Not applicable'' for studies not involving humans. You might also choose to exclude this statement if the study did not involve humans. Written informed consent for publication must be obtained from participating patients who can be identified (including by the patients themselves). Please state ``Written informed consent has been obtained from the patient(s) to publish this paper'' if applicable.}

\dataavailability{{The~Python implementation of the volatility models proposed in this paper is available in \url{https://github.com/EduardoRamosP/MultiTransformer} (accessed on \mbox{26 June 2021}).} }%In this section, please provide details regarding where data supporting reported results can be found, including links to publicly archived datasets analyzed or generated during the study. Please refer to suggested Data Availability Statements in section ``MDPI Research Data Policies'' at \url{https://www.mdpi.com/ethics}. You might choose to exclude this statement if the study did not report any data.}

%\acknowledgments{In this section you can acknowledge any support given which is not covered by the author contribution or funding sections. This may include administrative and technical support, or donations in kind (e.g., materials used for experiments).}

\conflictsofinterest{The authors declare that they have no conflict of interest regarding the publication of the research~article.}

%% Optional
%\sampleavailability{Samples of the compounds ... are available from the authors.}

%%%%%%%%%%%%%%%%%%%%%%%%%%%%%%%%%%%%%%%%%%
%% Only for journal Encyclopedia
%\entrylink{The Link to this entry published on the encyclopedia platform.}

%%%%%%%%%%%%%%%%%%%%%%%%%%%%%%%%%%%%%%%%%%
%% Optional
%\abbreviations{The following abbreviations are used in this manuscript:\\

%\noindent
%\begin{tabular}{@{}ll}
%MDPI & Multidisciplinary Digital Publishing Institute\\
%DOAJ & Directory of open access journals\\
%TLA & Three letter acronym\\
%LD & Linear dichroism
%\end{tabular}}

%%%%%%%%%%%%%%%%%%%%%%%%%%%%%%%%%%%%%%%%%%

%%%%%%%%%%%%%%%%%%%%%%%%%%%%%%%%%%%%%%%%%%
\end{paracol}
\reftitle{References}


%\externalbibliography{yes}
%\bibliography{paper1}

\end{document}